\documentclass[nohyper,12pt,a4paper]{JHEP3}
\usepackage{epsfig}
\usepackage{psfrag}
\usepackage{graphicx}
\usepackage{amsfonts,amsmath,amssymb}

\newcommand{\DIV}[1][\normalsize]{\,\mbox{div}\,}

\newcommand{\eqnref}[1]{(\ref{#1})}

\def\R{ {\mathbb R} }

\def\beq{\begin{equation}}
\def\eeq{\end{equation}}
\def\bea{\begin{eqnarray}}
\def\eea{\end{eqnarray}}

\title{Thermodynamics of Dyonic Lifshitz Black Holes}

\author{
T. Zingg$^{1),2)}$\\ \ \\
1) University of Iceland, Science Institute \\Dunhaga 3, IS-107 Reykjavik, Iceland\\\ \\
2) NORDITA, Roslagstullsbacken 23 \\SE-106 91 Stockholm, Sweden \\ \ \\
E-mail:\ \email{zingg@nordita.org}
}

\abstract{
Black holes with asymptotic anisotropic scaling are conjectured to be gravity duals of condensed matter system close to quantum critical points with non-trivial dynamical exponent $z$ at finite temperature.
A holographic renormalization procedure is presented that allows thermodynamic potentials to be defined for objects with both electric and magnetic charge in such a way that standard thermodynamic relations hold.
Black holes in asymptotic Lifshitz spacetimes can exhibit paramagnetic behavior at low temperature limit for certain values of the critical exponent $z$, whereas the behavior of AdS black holes is always diamagnetic.

}

\preprint{
}

\begin{document}


\section{Introduction}

Taking the AdS/CFT correspondence \cite{Maldacena:1997re} as a guideline, geometries with an anisotropic scaling have been presented in \cite{Kachru:2008yh} as candidates for gravitational duals for quantum critical condensed matter systems that are invariant under Lifshitz scaling
\beq
t \to \lambda^z t, \, x \to \lambda x	\; ,
\eeq
with a dynamical exponent $z>1$.
A detailed understanding of quantum critical metals poses an challenge in theoretical physics \cite{2005Natur.433..226C}.
Using gravitational duals to shed new light on these and other condensed matter problems continues to be an active field of research (for reviews see e.g. \cite{Hartnoll:2009sz,McGreevy:2009xe,Sachdev:2010ch}).
While the validity of this formalism is still a matter of debate, it is important to develop these holographic duals further in order to be able to test them against experimental results.
This paper provides a prescription for defining thermodynamic quantities for dyonic Lifshitz black branes that satisfy the expected thermodynamic relations.
Such a prescription allows to investigate duals of systems with anisotropic scaling in the presence a magnetic field - along the same lines as e.g. AdS/CFT duality has been used to obtain a holographic description of the Hall conductivity \cite{Hartnoll:2007ai}.

The paper is structured as follows.
In section \ref{sec:mod}, the effective action of the holographic model is introduced.
It consists of an Einstein-Hilbert term coupled to a Proca field and a $U(1)$ gauge field, which is known to give rise to charged Lifshitz black brane solutions.
Building on previous analysis \cite{Taylor:2008tg,Bertoldi:2009vn,Brynjolfsson:2009ct,Ross:2009ar}, the model with both electric and magnetic fields present is investigated with an emphasis on holographic renormalization and thermodynamics.
The focus will be on 3+1 dimensions, i.e. a strongly interacting field theory dual in 2+1 dimensions.
This setup is relevant for a holographic description of materials where the charge carriers are confined to layers orthogonal to a magnetic field.
Another benefit of working in this dimension is the duality between magnetic and electric fields, which simplifies the following analysis.

In section \ref{sec:ren}, counterterms are introduced such that the action and its functional derivatives are well-defined and finite on-shell.
This extends the analysis in \cite{Ross:2009ar}, where no gauge-field was considered, and allows to renormalize the action for values $z\geq 2$, which goes beyond the parameter range considered in \cite{Dehghani:2011tx}.
The prescription presented here is based on a new approach to identify the degrees of freedom of the system.
Within the space of solutions, Lifshitz-spacetimes form a subset which is disconnected from other classes of solutions.
Thus, on-shell variations must be constrained such that they do not
lead away from that subspace.

The renormalization procedure is then used in section \ref{sec:thermo} to define the internal energy and Helmholtz free energy of the black brane solutions.
These obey the same relations as in standard thermodynamics.
This extends the work of \cite{Bertoldi:2009vn,Dehghani:2011tx} and gives further justification that the thermodynamic description of black holes, which is known to be valid in the AdS-case, is also applicable for non-relativistic holography.

\section{The holographic model}
\label{sec:mod}


The effective action used is of the form introduced in \cite{Taylor:2008tg} with a Maxwell term added,
\bea
S_{bulk}	&=&	S_{Einstein} + S_{Proca} + S_{Maxwell}	\; .
\label{eq:action_bulk}
\eea
In the above,
\bea
S_{Einstein}	&=&	\frac{1}{2 \kappa L^2} \int_M (R-2\Lambda) \mathit{v}_M	\; ,	\label{eq:S_Einstein}	\\
S_{Proca}	&=&	-\frac{1}{4 \kappa L^2} \int_M \left( dP \wedge * dP + \mathit{c}P\wedge P\right)	\; ,	\label{eq:S_Proca}	\\
S_{Maxwell}	&=&	-\frac{1}{4 \kappa L^2} \int_M F \wedge F	\; ,	\label{eq:S_Maxwell}
\eea
where $P$ is a $1$-form on the manifold $M$ and $F=dA$ the Maxwell-tensor.
Variation with respect to the metric $g_{\mu\nu}$ gives the Einstein equations
\bea
G_{\mu\nu} + \Lambda g_{\mu\nu}	&=&	T^P_{\mu\nu} + T^{EM}_{\mu\nu}	\; .	\label{eq:Einstein}
\eea
The energy tensors $T^P_{\mu\nu}$ and $T^{EM}_{\mu\nu}$ are defined as
\bea
T^P_{\mu\nu}	&=&	\frac{1}{2}\left(	P_{\mu}P_{\nu}
						+ [dP]_{\mu\lambda}{[dP]_{\nu}}^{\lambda}
						-\frac{1}{2}P_{\lambda}P^{\lambda}
						-\frac{1}{4}[dP]_{\lambda\kappa}[dP]^{\lambda\kappa}
			\right)	\; ,	\label{eq:T_P}	\\
T^{EM}_{\mu\nu}	&=&	\frac{1}{2}\left(	F_{\mu\lambda}{F_{\nu}}^{\lambda}
						-\frac{1}{4}F_{\lambda\kappa}F^{\lambda\kappa}
			\right)	\; .	\label{eq:T_EM}
\eea
Variation with respect to $P$ and $A$ leads to the equations of motion
\bea
d*dP	&=&	-\mathit{c} *P	\; ,	\label{eq:eom_P}	\\
d*F	&=&	0	\; .	\label{eq:eom_F}
\eea
Equations \eqnref{eq:Einstein}, \eqnref{eq:eom_P} and \eqnref{eq:eom_F} are known to have asymptotic Lifshitz solutions with dynamical exponent $z$ if
\bea
\Lambda		&=&	-\frac{z^2+(d-1)z+d^2}{2 L^2}	\; ,	\label{eq:Lambda}	\\
\mathit{c}	&=&	\frac{d z}{L^2}	\; ,	\label{eq:c}
\eea
where $d$ is related to the dimension of $M$ by $d+2=\dim M$.

For concreteness, the case $\dim M = 4$, i.e. $d=2$, is used in the following.
This is partly motivated by previous investigations \cite{Brynjolfsson:2009ct,Brynjolfsson:2010rx,Brynjolfsson:2010mk}, where results so far indicated that the qualitative behavior of the system is mainly characterized by the ratio $d/z$ rather than $d$ itself.
Thus, using general $d$ would merely clutter notation without being much more instructive.
Beyond that, the duality between electric and magnetic field strength in $3+1$ dimensions allows for some further simplification.

\subsection{Dyonic black branes}
\label{sec:ansatz}

For the black branes, an ansatz of a static and stationary metric is chosen via the tetrad
\bea
e_0	&=&	L \frac{f g}{r^z}\, dt	\; ,	\label{eq:e_0}	\\
e_1	&=&	L \frac{1}{r}\, dx	\; ,	\label{eq:e_1}	\\
e_2	&=&	L \frac{1}{r}\, dy	\; ,	\label{eq:e_2}	\\
e_3	&=&	-L \frac{1}{r g}\, dr	\; .	\label{eq:e_3}
\eea
The metric is then given by $g_{\mu\nu}=\eta^{AB} e_{A\mu} e_{B\nu}$.
The tetrad has been introduced for later convenience.
Furthermore, the orientation on the manifold $M$ is chosen to be
\bea
\mathit{v}_M	&=&	e_3 \wedge e_1 \wedge e_2 \wedge e_0	\; .	\label{eq:v_M}
\eea
The Proca-field and gauge potential are parametrized as
\bea
P	&=&	\sqrt{\frac{2}{z}} L r^{-z} a f g^2 \,dt	\; ,	\label{eq:P}	\\
A	&=&	L r^{-z} \phi f g^2 \,dt + L B_0 x\,dy	\; .	\label{eq:A}
\eea
Here, $B_0$ is the constant field strength of a magnetic field perpendicular to the $xy$-plane.
For later convenience, $dP$ and $F$ are parametrized as
\bea
dP	&=&	-\sqrt{2z} L r^{-z-1} b f \,dr\wedge dt		\; ,	\label{eq:dP}	\\
F	&=&	L r^{-z+1} f \rho_0 \,dr\wedge dt + L B_0 \,dx\wedge dy	\; ,	\label{eq:F}
\eea
where the constant $\rho_0$ describes the charge density in the system.
The above defined ansatz will solve the equations of motion \eqnref{eq:Einstein}, \eqnref{eq:eom_P} and \eqnref{eq:eom_F} provided the following system of first order ODEs holds
\begin{eqnarray}
r f'
	&=&	f \left(z-1-a^2\right)	\; ,	\label{eq:bg_eqns_f}	\\
r g'
	&=&	\frac{g}{2} \left(3 + a^2\right) + \frac{1}{2 g}\left(\Lambda + \frac{\rho_0^2 + B_0^2}{4}r^4 + \frac{z}{2}b^2\right)	\; ,	\label{eq:bg_eqns_g}	\\
r a'
	&=&	-2 a - \frac{1}{g^2}\left[z b + a\left(\Lambda + \frac{\rho_0^2 + B_0^2}{4}r^4 + \frac{z}{2}b^2 \right)\right]	\; ,	\label{eq:bg_eqns_a}	\\
r b'
	&=&	2 b - 2 a	\; ,	\label{eq:bg_eqns_b}	\\
r \phi'
	&=&	-2 \phi + \frac{1}{g^2}\left[\rho_0 r^2 - \phi\left(\Lambda + \frac{\rho_0^2 + B_0^2}{4}r^4 + \frac{z}{2}b^2 \right)\right]	\; .	\label{eq:bg_eqns_phi}	
\end{eqnarray}
A straightforward calculation shows that the quantity
\bea
D_0	&=&	\frac{f}{r^{z+2}} \Bigl(
					-\Lambda
					-\frac{\rho_0^2+B_0^2}{4} r^4
					-\frac{1}{2} z b^2
					-3 g^2		\nonumber	\\
	& &	\qquad\qquad		+a^2 g^2
					-2 a b g^2
					+\frac{ \rho_0^2 + B_0^2}{\rho_0} r^2 g^2 \Phi
				\Bigr)
\label{eq:D_0}
\eea
is a first integral.
This quantity will prove useful in deriving an equation of state in section \ref{sec:thermo}.

The system \eqnref{eq:bg_eqns_f}-\eqnref{eq:bg_eqns_phi} has the asymptotic Lifshitz fixed point $f=f_\infty, g=1, a=b=\sqrt{z-1}, \phi=0$.
It is however not the only fixed point, as there would also be the possibility $f \sim r^{z-1}, g=\sqrt{\frac{4+z+z^2}{6}}, a=b=\phi=0$, which corresponds to an asymptotic AdS solution.
The system can be solved in the asymptotic region $r\to 0$ by first linearizing around the Lifshitz fixed point, which gives the asymptotic modes of the solution, and then iteratively calculating the descendants of these modes.
The expansion up to the orders that are relevant for this paper can be found in appendix \ref{app:exps}.
A crucial result of this expansion (up to a choice of sign) is
\bea
P	&=&	\left(\sqrt{\frac{2(z-1)}{z}} + \xi \right) e_0	\; ,	\label{eq:rel_P_e_0}
\eea
where $\xi$ is a scalar function that vanishes asymptotically for asymptotic Lifshitz solutions.
This will play an important role in the calculation of the on-shell variation in the next section.

\section{Renormalization}
\label{sec:ren}

Using the equations of motion \eqnref{eq:bg_eqns_f}-\eqnref{eq:bg_eqns_phi}, the bulk action can be shown to reduce to a surface term on-shell
\bea
S_{bulk}^{on-shell}	
	&=&	\frac{1}{2 \kappa L^3} \int_{\partial M} g \left( 2 -\frac{b_0^2 r^{2} \phi }{\rho_0} \right) \mathit{v}_{\partial M}	\; .
\label{eq:action_bulk_on-shell}
\eea
In the above expression, $\mathit{v}_{\partial M}$ is the induced volume form on the surface, i.e. $\mathit{v}_{\partial M} = e_1 \wedge e_2 \wedge e_0$.
Plugging in the expansion from appendix \ref{app:exps}
shows that the integrand diverges at the boundary as $r\to 0$.
Thus, the action needs to be renormalized before the standard thermodynamic gauge-gravity dictionary can be applied.
This can be done as for asymptotic AdS solutions \cite{Balasubramanian:1999re}, that is by confining the integration domain to $r>\varepsilon$ and defining a series of counterterms on the boundary $\partial M_\varepsilon$, such that the limit $\varepsilon\to 0$ is well-defined on shell.
It should be noted that the following analysis is also made under the condition $\Xi_1=0$, where $\Xi_1$ is a coefficient in the expansion from appendix \ref{app:exps}.
For $z\geq 2$ $(z\geq d)$ this would anyway be a necessary condition, as otherwise there would be a non-renormalizable mode in the solution, while for $z<2$ $(z<d)$, this choice greatly simplifies the definition of an energy, as it vastly reduces the number of required counterterms to cancel all divergent contributions\footnote{Similar conclusions were drawn in \cite{Ross:2009ar}.}.
Furthermore, it will also be assumed that $z<6$ $(z<d+4)$.
With this, it will be sufficient to consider counterterms which are at most quadratic in $F$ and $A$, which simplifies the analysis.
At the same time, most known quantum critical systems in experimental physics have a dynamical exponent in the range $1<z<3$, which is well within the range considered.
It can be expected, however, that the final result for the thermodynamic quantities in section \ref{sec:thermo}  remains the same for higher values of $z$, but
an explicit expression for the renormalized action would involve terms which are quartic\footnote{There won't be any terms with odd powers as the equations of motion are invariant under charge conjugation $A \to -A$.} and higher order in $F$ and $A$.
Finally, the normalization $f_\infty=1$ is used throughout.
This is not a real restriction, as another value would simply correspond to a different choice of the time scale, which could be easily be reintroduced by multiplying expressions with the appropriate power of $f_\infty$.

The final form of the action is thus given by
\bea
S	&=&	S_{bulk} + S_{reg} + S_{ct}	\; .
\label{eq:action_final}
\eea
$S_{reg}$ is a regulating term necessary to have a well defined variation in the presence of a boundary,
\bea
2\kappa L^2 S_{reg}	&=&	2\int_{\partial M_\varepsilon} K v_{\partial M_\varepsilon}
		-\int_{\partial M_\varepsilon} A \wedge * F
		-\frac{L}{(2-z)} \int_{\partial M} \mathfrak{L}_{e_3}A * F		\; .
\label{eq:action_reg}
\eea
The first term is the usual Gibbons-Hawking term which ensures that the variation with respect to the metric is well defined.
$K$ is the trace of the intrinsic curvature $K_{\mu\nu} = \nabla_{(\mu}n_{\nu)}$, which in the given setup can be calculated as $K_{\mu \nu} = \frac{g}{2L}\frac{\partial}{\partial r} h_{\mu\nu}$, where $h_{\mu\nu}=g_{\mu\nu}-n_\mu n_\nu$ is the induced metric on $\partial M_\varepsilon$.
The
second term
accounts for working in a background with fixed charged density, i.e. $\imath_{e_3} \delta F$, on the boundary instead of a fixed chemical potential.\footnote{This point is explained in e.g. \cite{McGreevy:2009xe}.}
This choice comes from the consideration that the holographic field theoretic problem in mind for this setup is a sample with $2$-dimensional (semi-)conducting layers in a constant magnetic field with a fixed number of dopants rather than a given chemical potential.
In the third term, $\mathfrak{L}_{e_3}$ is the Lie derivative with respect to $e_3$, which can be interpreted as the normal derivative on the boundary.
This last term is required to cancel a divergent contribution of the preceding term that occurs for $z>2$.
Introducing $\star$ to denote the induced Hodge-Star on the boundary and
decomposing the Maxwell tensor as
\begin{eqnarray}
F^e	&=&	\imath_{e_1} F			\; ,	\label{eq:F^e}	\\
F^m	&=&	F\Bigl|_{\partial M}	\; ,	\label{eq:F^m}
\end{eqnarray}
the regulating term \eqnref{eq:action_reg} can also be written as
\bea
2\kappa L^2 S_{reg}	&=&	2\int_{\partial M_\varepsilon} K v_{\partial M_\varepsilon}
		-\int_{\partial M_\varepsilon} A \wedge \star F^e
		-\frac{L}{(2-z)} \int_{\partial M} \mathfrak{L}_{e_3}A \wedge \star F^e		\; .
\label{eq:action_reg_2}
\eea
This makes it more manifest that the last two terms correspond to a Legendre transformation.
In the standard dictionary, a bulk gauge field $A$ will source a current $\mathcal{J}$ in the dual field theory on the boundary.
In that theory, $F^e$ has an interpretation as the response, i.e. $\langle \mathcal{J} \rangle \sim \frac{\delta S}{\delta A} \sim F^e$.
For $z>2$, it can be read off from \eqnref{eq:phi_asymptotics} that the mode with $\mathcal{Q}\nu$ will grow faster than the mode involving $\mathcal{Q}$.
Thus, the latter needs to be canceled and it is actually $A+\frac{L}{(2-z)} \mathfrak{L}_{e_3}A$ that becomes the source for the dual current.\footnote{This works analogous to the discussion in \cite{Hartnoll:2009ns}.}
The Legendre transformation then interchanges the role of $A+\frac{L}{(2-z)} \mathfrak{L}_{e_3}A$ as source and $F^e$ as response.

Finally, $S_{ct}$ is a counterterm that cancels all remaining divergences from the bulk action.
It is given by
\bea
2\kappa L^2 S_{ct}	&=&	\frac{2(z+1)}{L} \int_{\partial M} v_{\partial M}
		+\frac{\sqrt{2z(z-1)}}{L} \int_{\partial M_\varepsilon} \xi v_{\partial M}	\nonumber	\\
	& &	+\frac{L}{2 (2-z)} \int_{\partial M} \mathfrak{L}_{e_3}A \wedge \star \mathfrak{L}_{e_3}A
		+\frac{L}{2 (2-z)} \int_{\partial M} F^m \wedge \star F^m	\; ,
\label{eq:action_counterterm}
\eea
where $\xi$ was introduced in \eqnref{eq:rel_P_e_0}.
The first term in \eqnref{eq:action_counterterm} is simply a boundary cosmological constant.
The second term cancels a divergence with exponent $z_2$ coming from $\Xi_2$ having a non-vanishing value (cf. appendix \ref{app:exps}).\footnote{It is worth noting that the $\xi$-term also cancels the divergence proportional to $z_1$ when $\Xi_1\neq 0$. To renormalize the action, however, terms with higher powers in $\xi$ would also need to be included.}
For $1<z\leq 2$ these two terms would actually suffice, for $z>2$ there are however further divergences occurring due to the electric and magnetic fields not falling off fast enough.
This is cured by adding the terms in the second line of \eqnref{eq:action_counterterm} involving $\mathfrak{L}_{e_3}A$ and $F^m$.

Two issues might need some clarification.
First, the attentive reader may have noticed that there is a certain redundancy in the notation as $\mathfrak{L}_{e_3}A=F^e$.
This is done on purpose to make the conceptual difference between those two terms manifest.
$\mathfrak{L}_{e_3}A$ is defined through $A$, which is the field that enters the bulk action \eqnref{eq:action_bulk}.
The field $F^e$ is introduced by performing a Legendre transformation.
In the logical order, this transformation is done after the action was renormalized.
Therefore, the counterterm \eqnref{eq:action_counterterm} is written in terms of $\mathfrak{L}_{e_3}A$ and without any explicit dependence on $F^e$.

Second, the first two terms in \eqnref{eq:action_counterterm} which cancel all divergent contributions when $A=0$ are different from the terms proposed in \cite{Ross:2009ar}, where
\bea
2\kappa L^2 \tilde{S}_{ct}	&=&	\frac{4}{L} \int_{\partial M} v_{\partial M}
		+\frac{\sqrt{2z(z-1)}}{L} \int_{\partial M_\varepsilon} \sqrt{\langle P, P \rangle} v_{\partial M}	\; .
\label{eq:action_counterterm_Ross_Saremi}
\eea
However, by using \eqnref{eq:rel_P_e_0}, a short calculation reveals
\bea
4 + \sqrt{2z(z-1)}\sqrt{\langle P, P \rangle}
	&=&	(z+3) - \frac{z}{2} \langle P, P \rangle + O(\xi^2)	\nonumber	\\
	&=&	2(z+1) + \sqrt{2z(z-1)} \xi + O(\xi^2)	\; .
\eea
Thus, when $\xi^2 v_{\partial M}$ will vanish for $r\to 0$, which is indeed the case in the parameter range considered here, the first line of \eqnref{eq:action_counterterm} would give exactly the same contribution as \eqnref{eq:action_counterterm_Ross_Saremi}.

Energy and momentum can also be calculated along the lines of \cite{Balasubramanian:1999re}.
The procedure differs, however, in the following two ways.
First, as was already pointed out in \cite{Ross:2009ar}, the dual theory is not relativistic and thus it is less convenient to work with the metric and the stress energy tensor $T^{\mu\nu} = \frac{2}{\sqrt{-h}}\frac{\delta S}{\delta  h_{\mu\nu}}$, but more useful to work with a tetrad and $\tau_A$, where
\beq
\star\tau_A = \eta_{AB}\frac{\delta S}{\delta  e_B}	\; .
\label{eq:tau_def}
\eeq
Energy, momentum, energy flux and stress are then encoded in the components of $\tau_A$.
A second, and more subtle difference is the way the variations are calculated.
The aim is to consider gravity duals to systems with anisotropic scaling, but, as was noted earlier, in addition to the Lifshitz fixed point there also exists an AdS fixed point.
In fact, the asymptotic Lifshitz spacetimes form an isolated subset of the space of solutions that is disconnected from the subset of asymptotic AdS spacetimes.
Thus, in the same fashion as the covariant derivative on a surface embedded in $\R^N$ is basically
the derivative of the embedding space constrained to be evaluated on curves that do not lead away from the surface,
the variations must be constrained to 'curves' that stay inside the subspace of asymptotic Lifshitz solutions.
These 'curves' are defined by \eqnref{eq:rel_P_e_0}, which is a direct consequence of making an ansatz that has asymptotic anisotropic scaling.
Hence, the variation must be performed under the constraint that it is not $P$, but the scalar $\xi$ which is a degree of freedom of the system.\footnote{Making this statement simply based on the expansion \eqnref{eq:f_asymptotics}-\eqnref{eq:phi_asymptotics} might appear ad hoc.
For the purpose of this paper it could just be thought of as a mere working assumption, but an investigation of the PDE-system \eqnref{eq:Einstein}, \eqnref{eq:eom_P} and \eqnref{eq:eom_F} via a Fefferman-Graham like expansion (cf. \cite{FeffermanGraham}) reveals that the actual degrees of freedom are not the components of $P$ but are defined via projections to $e_0, e_1, e_2$.
These more general results will be reported on elsewhere \cite{TBA}.
}
Using this relation, \eqnref{eq:tau_def} becomes
\bea
2\kappa L^2 \tau_0	&=&	2K\cdot e_0
			-2 K e_0
			+ \left( \sqrt{\frac{2(z-1)}{z}} + \xi \right) \imath_{e_3} dP	\nonumber	\\
		& &	- \left( \langle A,F^e \rangle e_0 - \langle A,e_0 \rangle F^e -\langle F^e,e_0 \rangle A \right)	\nonumber	\\
		& &	- \frac{L}{2-z} \left( \langle \mathfrak{L}_{e_3}A,F^e \rangle e_0 - \langle \mathfrak{L}_{e_3}A,e_0 \rangle F^e -\langle F^e,e_0 \rangle \mathfrak{L}_{e_3}A \right)	\nonumber	\\
		& &	+ \frac{2(z+1)}{L} e_0
			+ \frac{\sqrt{2z(z-1)}}{L} \xi e_0	\nonumber	\\
		& &	+ \frac{L}{2(2-z)} \left( \langle \mathfrak{L}_{e_3}A,\mathfrak{L}_{e_3}A \rangle e_0 - 2 \langle \mathfrak{L}_{e_3}A,e_0 \rangle \mathfrak{L}_{e_3}A  \right)	\nonumber	\\
		& &	+ \frac{L}{2(2-z)} \left( \langle F^m,F^m \rangle e_0 + 2 F^m\cdot F^m \cdot e_0 \right)	\; .
\label{eq:t_0}
\eea
In the expression above, $M\cdot \omega$ denotes the contraction ${M_\mu}^\nu\omega_\nu$ for a $2$-tensor $M$ and a $1$-form $\omega$.
$\tau_1$ and $\tau_2$ are given by similar expressions, but without the terms involving $\xi$.
However, only $\tau_0$ will be relevant when defining the energy in the next section.

For when considering differentials of thermodynamic quantities later on, it is also useful to note the relation
\beq
2\kappa L^2 \frac{\delta S}{\delta F^e} = -\star \left[ A + \frac{L}{2-z} \mathfrak{L}_{e_3}A \right]	\; .
\label{eq:dSdF}
\eeq

\subsection{A note about $z=2$}

For $z=2$, and more generally $z=d$, the asymptotic expansions (see appendix \ref{app:exps}) become anomalous and contain logarithmic terms.
Furthermore \eqnref{eq:action_reg_2}, \eqnref{eq:action_counterterm} \eqnref{eq:t_0}, and \eqnref{eq:dSdF} are not well-defined due to the factor of $(2-z)$ in the denominator.
In this special case these need to be modified,
\bea
2\kappa L^2 S_{reg}	&=&	2\int_{\partial M_\varepsilon} K v_{\partial M_\varepsilon}
		-\int_{\partial M_\varepsilon} A \wedge \star F^e
		-L \int_{\partial M} \ln r\, \mathfrak{L}_{e_3}A \wedge \star F^e		\; ,
\label{eq:action_reg_z=2}
\eea
\bea
2\kappa L^2 S_{ct}	&=&	\frac{2(z+1)}{L} \int_{\partial M} v_{\partial M}
		+\frac{\sqrt{2z(z-1)}}{L} \int_{\partial M_\varepsilon} \xi v_{\partial M}	\nonumber	\\
	& &	+\frac{L}{2} \int_{\partial M} \ln r\, \mathfrak{L}_{e_3}A \wedge \star \mathfrak{L}_{e_3}A
		+\frac{L}{2} \int_{\partial M} \ln r\, F^m \wedge \star F^m	\; ,	
\label{eq:action_counterterm_z=2}
\eea
\bea
2\kappa L^2 \tau_0	&=&	2K\cdot e_0
			-2 K e_0
			- \left( \sqrt{\frac{2(z-1)}{z}} + \xi \right) \imath_{e_3} dP	\nonumber	\\
		& &	- \left( \langle A,F^e \rangle e_0 - \langle A,e_0 \rangle F^e -\langle F^e,e_0 \rangle A \right)	\nonumber	\\
		& &	- L \ln r\, \left( \langle \mathfrak{L}_{e_3}A,F^e \rangle e_0 - \langle \mathfrak{L}_{e_3}A,e_0 \rangle F^e -\langle F^e,e_0 \rangle \mathfrak{L}_{e_3}A \right)	\nonumber	\\
		& &	+ \frac{2(z+1)}{L} e_0
			+ \frac{\sqrt{2z(z-1)}}{L} \xi e_0	\nonumber	\\
		& &	+ \frac{L}{2} \ln r\, \left( \langle \mathfrak{L}_{e_3}A ,\mathfrak{L}_{e_3}A  \rangle e_0 - 2 \langle \mathfrak{L}_{e_3}A ,e_0 \rangle \mathfrak{L}_{e_3}A   \right)	\nonumber	\\
		& &	+ \frac{L}{2} \ln r\, \left( \langle F^m,F^m \rangle e_0 + 2 F^m\cdot F^m \cdot e_0 \right)	\; ,
\label{eq:t_0_z=2}
\eea
\beq
2\kappa L^2 \frac{\delta S}{\delta F^e} = -\star \left[ A + L \ln r\, \mathfrak{L}_{e_3}A \right]	\; .
\label{eq:dSdF_z=2}
\eeq

\section{Thermodynamics}
\label{sec:thermo}


A temperature is introduced in the dual theory by considering black brane solutions with an event horizon at some finite value $r=r_0$.
At the horizon, $f\to f_0,a\to a_0,b\to b_0,\phi\to \phi_0$ while $g^2\to g_0^2 (1-r/r_0)$.
As the form of the equations is invariant under the rescaling $r\to\lambda r, B_0 \to \lambda^{-2}B_0, \rho_0 \to \lambda^{-2}\rho_0$, the horizon can be assumed to be at $r_0=1$.
The dependence of the solutions on $r_0$ can then be introduced by using the rescaling backwards.
With this simplification, the relation between the constants at the horizon is given by
\bea
g_0^2	&=&	-\Lambda-\frac{z}{2} b_0^2 -\frac{\rho_0^2+B_0^2}{4}	\; ,	\label{eq:g_0}	\\
a_0	&=&	\frac{z b_0}{g_0^2}		\; ,	\label{eq:a_0}	\\
\phi_0	&=&	-\frac{\rho_0}{g_0^2}	\; .	\label{eq:rho_0}
\eea
Thermodynamic quantities can now be assigned using the same prescription as for the $AdS$ case (see e.g. \cite{Hartnoll:2009sz}).
The value of $\kappa$ is associated with the number of flavors $N$ in the dual field theory through $\frac{1}{\kappa}=\frac{\sqrt{2}N^{\frac{3}{2}}}{3 \pi}$.
The chemical potential $\mu$, magnetic field strength $\mathfrak{b}$ and the charge density $\mathfrak{q}$ can be read off from the asymptotic expansion in appendix \ref{app:exps},
\beq
\mu		=	\frac{\mathcal{Q} \nu}{L}	\; ,
\mathfrak{b}	=	\frac{\mathcal{B}}{L^2}	\; ,
\mathfrak{q}	=	\frac{\mathcal{Q}}{2\kappa L^2}		= \frac{\sqrt{2}N^{\frac{3}{2}} \mathcal{Q}}{6\pi L^2} \; .
\eeq
As $B_0$ and $\rho_0$ enter in a symmetric fashion in the equations of motion \eqnref{eq:bg_eqns_f}-\eqnref{eq:bg_eqns_phi}, the magnetization density is given by
\beq
\mathfrak{m}	=	- \frac{\mathcal{B} \nu}{2 \kappa L}	=	-\frac{1}{4\kappa^2}\frac{\mathfrak{b}\mu}{\mathfrak{q}}	\; .
\label{eg:magnetisation}
\eeq
This relation will become more clear in the discussion in subsection \ref{sec:diff}.
Reintroducing the scaling in $r_0$ reveals that $\mathfrak{m},\mu \propto r_0^{-z}$
and the values of $\mathfrak{b}$ and $\mathfrak{q}$
are related to the variables at the horizon via
\beq
\mathfrak{b}	=	\frac{B_0}{L^2 r_0^2}	\; ,
\mathfrak{q}	=	\frac{\rho_0}{2\kappa L^2 r_0^2}	\, .
\eeq
A temperature is defined via Wick rotating time
and then compactifying on the thermal circle, the result is
\beq
T	=	\frac{f_0 g_0^2}{4 \pi r_0^z L}	\; .
\label{eq:T_DLBH}
\eeq
The value of $r_0$ also defines the entropy density,
\beq
\mathfrak{s}	=	\frac{2 \pi}{\kappa L^2 r_0^2}	\; .
\label{eq:Entropy}
\eeq
The evaluation of the conserved quantity \eqnref{eq:D_0} at the horizon and in the region $r\to 0$ relates $T$ and $\mathfrak{s}$ with the variables in appendix \ref{app:exps}.
\beq
2\kappa L^3 \mathfrak{s} T
	=\left\{\begin{array}{lll}
			-\frac{2 \sqrt{z-1} (z^2-4) \mathcal{M}}{z}-(\mathcal{B}^2+\mathcal{Q}^2)\frac{ \mu }{\mathcal{Q}}		& \quad	&	z\neq 2	\; ,	\\
			-4\mathcal{M} -\frac{1}{4}(\mathcal{B}^2+\mathcal{Q}^2)-(\mathcal{B}^2+\mathcal{Q}^2)\frac{ \mu }{\mathcal{Q}}	& \quad	&	z=2	\; .
		\end{array}\right.
\label{eq:EOM_0}
\eeq

\subsection{Thermodynamic potential and equation of state}

By the standard prescription, the grand canonical potential is associated with the the value of the renormalized Euclidean on-shell action.
However, in the case at hand \eqnref{eq:action_final} contains the term $-\int_{\partial M_\varepsilon} \left( A + \frac{L}{(2-z)} \mathfrak{L}_{e_3}A \right) \wedge \star F^e$ which has been added to the action to allow for a setup with fixed charge density.
This terms is not part of the renormalization to cancel divergences, but it changes the thermodynamic potential by the value $\mu \mathfrak{q}$, resulting in the canonical ensemble.
Thus, from the on shell value of the action \eqnref{eq:action_final},
\beq
\mathfrak{a} \mathcal{V}	= T S^{Eucl,on-shell}	\;	.
\label{eq:Helmholtz_def}
\eeq
$\mathcal{V}$ is the volume of the system and $\mathfrak{a}$ is the Helmholtz free energy density.
Plugging in the parametrization presented in section \ref{sec:ansatz} and 
using the asymptotic expansion from appendix \ref{app:exps} leads to
\beq
2\kappa L^3 \mathfrak{a} = 
	\left\{\begin{array}{lll}
			2 (z-2) \sqrt{z-1} \mathcal{M}+\frac{\mathcal{B}^2 \mu }{\mathcal{Q}}+\mathcal{Q}  \mu				& \quad	&	z\neq 2	\; ,	\\
			2\mathcal{M} +\frac{1}{4}(\mathcal{B}^2+\mathcal{Q}^2)+\frac{\mathcal{B}^2 \mu }{\mathcal{Q}}+\mathcal{Q}\mu 	& \quad	&	z=2	\; .
		\end{array}\right.
\label{eq:Helmholtz_FE}
\eeq
An internal energy can be defined by working in the spirit of the AdS/CFT correspondence and considering the on-shell action \eqnref{eq:action_final} as a generating functional for the dual field theory with the boundary values of the fields interpreted as sources for their dual operators.
In \cite{Balasubramanian:1999re} the dual stress energy tensor $T^{\mu\nu}$ was considered as the operator that is sourced by the boundary metric $h_{\mu\nu}$.
As already indicated in section \ref{sec:ren}, instead of $T^{\mu\nu}$, the quantities $\tau_A$ defined in \eqnref{eq:tau_def} will now be used for this purpose.
With $\partial_t$ being the Killing vector that generates time translation invariance
the internal energy density $\mathfrak{e}$ is associated to $\tau_0$, given in \eqnref{eq:t_0}, through
\beq
L\mathfrak{e} = \tau_0(\partial_t) \Bigr|_{sources=0}	\; .
\label{eq:e_def}
\eeq
The subscript $sources=0$ reminds of the fact that according to the standard description, the right hand side of \eqnref{eq:e_def}, which comes from a functional derivative of the on-shell action, must be evaluated with all sources, i.e. independent boundary values, set equal to zero.
To account for this, $\tau_0(\partial_t)$ must be evaluated at the point where the explicit dependence on the source $F^e$ is set equal to zero.
Formulated quantitatively, this means that the term $-\int_{\partial M_\varepsilon} \left( A + \frac{L}{(2-z)} \mathfrak{L}_{e_3}A \right) \wedge \star F^e$ in \eqnref{eq:action_reg_2} will not contribute to the internal energy.
This is also sensible, as this term would give a contribution coming from having a nonzero chemical potential in the system, whereas the internal energy by definition should just account for the mass of the black brane that causes the curvature of spacetime.
The result is 
\beq
2\kappa L^3 \mathfrak{e} = 
	\left\{\begin{array}{lll}
			-\frac{4 (z-2) \sqrt{z-1} \mathcal{M}}{z}	& \quad	&	z\neq 2	\; ,	\\
			-2\mathcal{M}					& \quad	&	z=2	\; .
		\end{array}\right.
\label{eq:internal_E}
\eeq
Now \eqnref{eq:EOM_0}, \eqnref{eq:Helmholtz_FE} and \eqnref{eq:internal_E} can be combined to
\beq
\mathfrak{a}	=	\mathfrak{e} - \mathfrak{s}T 	\; ,
\label{eq:thermo_rel_1}
\eeq
which is indeed the correct expression for the density of the Helmholtz free energy.
Furthermore, from the above calculations an equation of state can be derived.
\beq
\frac{z+2}{2}\mathfrak{e} = 
	\left\{\begin{array}{lll}
			\mathfrak{s}T - \mathfrak{m} \mathfrak{b} + \mu \mathfrak{q}				& \quad	&	z\neq 2	\; ,	\\
			\mathfrak{s}T - \mathfrak{m} \mathfrak{b} + \mu \mathfrak{q}
					 +\frac{L}{8\kappa}\mathfrak{b}^2 + \frac{\kappa L}{2} \mathfrak{q}^2	&\quad	&	z=2	\; .
		\end{array}\right.
\label{eq:EOS}
\eeq
The first line is in accord with the findings in \cite{Bertoldi:2009dt} and \cite{Dehghani:2011tx}.
The appearance of $\mathfrak{b}^2$ and $\mathfrak{q}^2$ in the equation of state for $z=2$ is an artifact of an ambiguity in defining a counterterm for this particular value of the dynamical critical exponent.
The approach presented in section \ref{sec:ren} was a minimal one, i.e. just taking the counterterms which are required to cancel all divergences.
This results in the coefficients in \eqnref{eq:EOS}.
As a matter of fact, for $z=2$ it would be possible to add the terms $2\int_{\partial M} \mathfrak{L}_{e_3} A \wedge \star F^e - \int_{\partial M} \mathfrak{L}_{e_3} A \wedge \star \mathfrak{L}_{e_3} A$ and $\int_{\partial M} F^m \wedge \star F^m$
with arbitrary coefficients to the action.
This would leave \eqnref{eq:thermo_rel_1} unchanged, but would alter the coefficients of $\mathfrak{b}^2$ and $\mathfrak{q}^2$ in \eqnref{eq:EOS}.
In particular, it would be possible to cancel these coefficients, making the second line of \eqnref{eq:EOS} identical to the first.
It is unclear, what argument should be used to single out this choice and fix the ambiguity.

\subsection{The differential of the Helmholtz free energy}
\label{sec:diff}

In thermodynamics, $\mathfrak{a}$ satisfies
\beq
d\mathfrak{a}	=	- \mathfrak{s}\,dT - \mathfrak{m}\,d\mathfrak{b} + \mu\,d\mathfrak{q}	\; .
\label{eq:d_Helmholtz}
\eeq
This corresponds to the three relations
\bea
\left.\frac{\partial\mathfrak{a}}{\partial T}\right|_{\mathfrak{b},\mathfrak{q}}	& = &	- \mathfrak{s}	\; ,	\label{eq:daT}	\\
\left.\frac{\partial\mathfrak{a}}{\partial \mathfrak{b}}\right|_{T,\mathfrak{q}}	& = &	- \mathfrak{m}	\; ,	\label{eq:dab}	\\
\left.\frac{\partial\mathfrak{a}}{\partial \mathfrak{q}}\right|_{T,\mathfrak{b}}	& = &	\mu		\; .	\label{eq:daq}
\eea
These can easily be verified for dyonic AdS black branes in the case of $z=1$, where an exact solution is known.
What will be shown in the following is that they also hold for $z>1$.

First of all, \eqnref{eq:daq} is a direct consequence of \eqnref{eq:dSdF} when taking the limit $r\to 0$.
From this, \eqnref{eq:d_Helmholtz} will follow if it can be shown that any of the relations \eqnref{eq:daT}-\eqnref{eq:daq} implies the other two.
To proceed with the proof of this, it is useful to note that in the equations of motion \eqnref{eq:bg_eqns_f}-\eqnref{eq:bg_eqns_phi} as well as \eqnref{eq:Helmholtz_FE},
the values of $B_0$ and $\rho_0$ only occur in the combination
\beq
\eta = B_0^2+\rho_0^2	\; .
\label{eq:eta}
\eeq
Furthermore, as the dependence of $r_0$ just enters in the form of a rescaling of the final expression, all so far introduced thermodynamic quantities must be of the form $\frac{\Omega(\eta)}{r_0^s}$ with $s$ being some scaling exponent and $\Omega$ a function of a single variable.\footnote{Of course, $\mathfrak{q}$, $\mu$, $\mathfrak{b}$ and $\mathfrak{m}$ are not exactly of this form, they however differ only by a factor of $\mathfrak{q}$ or $\mathfrak{b}$ respectively.}
Therefore, let the functions $\mathcal{F}$ and $\mathcal{G}$ be defined via
\bea
T	&=& \frac{\mathcal{F}(\eta)}{4 \pi L r_0^z}	\; ,	\label{eq:def_F}	\\
\frac{1}{2\kappa} \frac{\mu}{\mathfrak{q}}	&=& -2\kappa\frac{\mathfrak{m}}{\mathfrak{b}} =	\frac{L \mathcal{G}(\eta)}{r_0^{z-2}}	\; .	\label{eq:def_G}
\eea
Imposing the conditions $r_0=1$ and $\Xi_1=0$ on the ODE system \eqnref{eq:bg_eqns_f}-\eqnref{eq:bg_eqns_phi} results in a one-parameter family of solutions, the parameter being $\eta$.
Hence, $\mathcal{F}$ and $\mathcal{G}$ are not independent and must satisfy a non-trivial relation.
This relation turns out to be
\beq
4\mathcal{F}' - (z-2)\mathcal{G} + 4 \eta \mathcal{G}'	= 0	\; .
\label{eq:FGcond}
\eeq
The validity of this will follow as a corollary to what will be proved in the following, namely that \eqnref{eq:FGcond} is equivalent to each of \eqnref{eq:daT}-\eqnref{eq:daq}.

First of all, the differentials of $T$, $\mathfrak{b}$ and $\mathfrak{q}$ are
\bea
dT		&=&	-\frac{z \mathcal{F}}{4 \pi L r_0^{z+1}} dr_0 + \frac{2 \mathcal{F}'}{4 \pi L r_0^{z}} (B_0\,dB_0 + \rho_0\,d\rho_0 )	\; ,	\\
d\mathfrak{b}	&=&	-\frac{2 B_0}{L^2 r_0^{3}} dr_0 + \frac{1}{L^2 r_0^{2}} dB_0	\; ,	\\
d\mathfrak{q}	&=&	-\frac{\rho_0}{\kappa L^2 r_0^{3}} dr_0 + \frac{1}{2\kappa L^2 r_0^{2}} d\rho_0	\; .
\eea
From this, at constant $\mathfrak{b}$ and $\mathfrak{q}$,
\bea
dT\Bigr|_{\mathfrak{b},\mathfrak{q}}
	&=&	\frac{1}{4 \pi L r_0^{z+1}}\left(-z \mathcal{F} + 4 \eta \mathcal{F}' \right) \,dr_0	\; ,	\\
d(\mathfrak{s}T)\Bigr|_{\mathfrak{b},\mathfrak{q}}
	&=&	\frac{1}{2\kappa L^3 r_0^{z+3}}\left[-(z+2) \mathcal{F} + 4 \eta \mathcal{F}' \right] \,dr_0	\; ,	\\
\left.d\left(\frac{\mu}{\mathfrak{q}}\right)\right|_{\mathfrak{b},\mathfrak{q}}
	&=&	\frac{2\kappa L}{r_0^{z-1}}\left[-(z-2) \mathcal{G} + 4 \eta \mathcal{G}' \right] \,dr_0	\; .
\eea
Therefore,
\bea
\left.\frac{\partial\mathfrak{a}}{\partial T}\right|_{\mathfrak{b},\mathfrak{q}}
	&=&	\left.\frac{\partial}{\partial T}\left[-\frac{z}{z+2} T \mathfrak{s} + \frac{2}{z+2} \frac{\mathfrak{b}^2+4\kappa^2\mathfrak{q}^2}{4\kappa^2}\frac{\mu}{\mathfrak{q}} \right]\right|_{\mathfrak{b},\mathfrak{q}}	\nonumber	\\
	&=&	-\frac{2 \pi}{\kappa L^2 r_0^2\left(z \mathcal{F}-4\eta \mathcal{F}'\right)}\left[ z \mathcal{F} -\frac{4 z}{z+2}\eta \mathcal{F}' - \frac{2 (z-2)}{z+2}\eta \mathcal{G} + \frac{8}{z+2}\eta^2 \mathcal{G}' \right]	\nonumber	\\
	&=&	-\frac{2 \pi}{\kappa L^2 r_0^2} - \frac{4 \pi \eta [4 \mathcal{F}' - (z-2)\mathcal{G} + 4\eta \mathcal{G}']}{(z+2) \kappa L^2 r_0^2\left(z \mathcal{F}-4\eta \mathcal{F}'\right)}
	\nonumber	\\
	&=&	-\mathfrak{s}	\; .
\eea
The last equality follows from \eqnref{eq:FGcond}.
This establishes the equivalence of \eqnref{eq:FGcond} and \eqnref{eq:daT}.

In an analogous way for constant $T$ and $\mathfrak{q}$,
\bea
d\mathfrak{b}\Bigr|_{T,\mathfrak{q}}
	&=&	\frac{1}{2 L^2 B_0 r_0^3 \mathcal{F}'} \left(z \mathcal{F} - 4 \eta \mathcal{F}' \right) \,dr_0	\; ,	\\
d\mathfrak{s}\Bigr|_{T,\mathfrak{q}}
	&=&	-\frac{4 \pi }{\kappa L^2r_0^{3}}\,dr_0	\; ,	\\
\left.d\left(\frac{\mu}{\mathfrak{q}}\right)\right|_{T,\mathfrak{q}}
	&=&	\frac{2\kappa L}{r_0^{z-1} \mathcal{F}'}\left[-(z-2) \mathcal{F}' \mathcal{G} + z \mathcal{F} \mathcal{G}' \right] \,dr_0		\nonumber	\\
	&=&	\frac{\kappa L}{2 r_0^{z-1} \eta \mathcal{F}'}\left[(z-2)(z \mathcal{F} - 4 \eta \mathcal{F}') \mathcal{G} - 4 z \mathcal{F} \mathcal{F}' \right] \,dr_0	\; .
\eea
In the last line, \eqnref{eq:FGcond} was inserted.
With this,
\bea
\left.\frac{\partial\mathfrak{a}}{\partial \mathfrak{b}}\right|_{T,\mathfrak{q}}
	&=&	\left.\frac{\partial}{\partial \mathfrak{b}}\left[-\frac{z}{z+2} T \mathfrak{s} + \frac{2}{z+2} \frac{\mathfrak{b}^2+4\kappa^2\mathfrak{q}^2}{4\kappa^2}\frac{\mu}{\mathfrak{q}} \right]\right|_{T,\mathfrak{q}}	\nonumber	\\
	&=&	\left.-\frac{z}{z+2} T \frac{\partial \mathfrak{s}}{\partial \mathfrak{b}}\right|_{T,\mathfrak{q}}
		+ \frac{1}{z+2} \frac{\mathfrak{b} \mu}{\kappa^2\mathfrak{q}}
		+ \left.\frac{2}{z+2} \frac{\mathfrak{b}^2+4\kappa^2\mathfrak{q}^2}{4\kappa^2}\frac{\mu}{\mathfrak{q}}\frac{\partial }{\partial \mathfrak{b}}\left(\frac{\mu}{\mathfrak{q}}\right)
		\right|_{T,\mathfrak{q}}	\nonumber	\\
	&=&	\frac{1}{z+2} \frac{\mathfrak{b} \mu}{\kappa^2\mathfrak{q}}
		+\frac{z-2}{z+2} \frac{L B_0 (\mathfrak{b}^2+4\kappa^2\mathfrak{q}^2) }{ 4\kappa r_0^{z-4}\eta}\mathcal{G}	\nonumber	\\
	& &	\quad	+\frac{B_0 r_0^3}{L\kappa\left(z \mathcal{F}-4\eta \mathcal{F}'\right)}\left[ 
									\frac{2 z}{z+2}\frac{\mathcal{F} \mathcal{F}'}{r_0^{z+3}}
									-\frac{2 z}{z+2}\frac{L^4 (\mathfrak{b}^2+4\kappa^2\mathfrak{q}^2) \mathcal{F} \mathcal{F}'}{r_0^{z-1}\eta}
								\right]	\nonumber	\\
	&=&	\frac{\mathfrak{b} \mu}{4\kappa^2 \mathfrak{q}}	\nonumber	\\
	&=&	-\mathfrak{m}	\; .
\eea
Again, as the equality holds if and only if \eqnref{eq:FGcond} is assumed, the equivalence of that assumption to \eqnref{eq:dab} is proved.
Due to the symmetric appearance of $\mathfrak{b}$ and $\mathfrak{q}$, this must also be true for \eqnref{eq:daq}.
As the validity of \eqnref{eq:daq} has already been established, this concludes the proof of \eqnref{eq:d_Helmholtz}.

Unfortunately, the thermodynamic relations found so far are not sufficient to determine an explicit expression for $\mathcal{F}$ or $\mathcal{G}$.
A few exact solutions are known (see appendix \ref{app:exact_solutions}), but in general numerical methods are needed to study these functions.

\subsection{Susceptibility and magnetization}

It is also possible to derive an expression for the density of the magnetic susceptibility,
\bea
\chi	&=&	\left. \frac{\partial \mathfrak{m}}{\partial \mathfrak{b}}\right|_{T,\mathfrak{q}}	\nonumber	\\
	&=&	\frac{\mathfrak{m}}{\mathfrak{b}}
		+\left. \mathfrak{b} \frac{\partial }{\partial \mathfrak{b}}\left(\frac{\mathfrak{m}}{\mathfrak{b}}\right)\right|_{T,\mathfrak{q}}	\nonumber	\\
	&=&	\frac{\mathfrak{m}}{\mathfrak{b}}
		- \mathfrak{b}\frac{L^3 B_0}{4\kappa r_0^{z-4}\eta}\left[(z-2)\mathcal{G} - \frac{4z\mathcal{F}\mathcal{F}'}{z \mathcal{F}-4\eta \mathcal{F}'}\right]	\nonumber	\\
	&=&	\frac{1}{2(\mathfrak{b}^2+4\kappa^2\mathfrak{q}^2)}\left[
			(z\mathfrak{b}^2+8\kappa^2\mathfrak{q}^2)\frac{\mathfrak{m}}{\mathfrak{b}}
			-\frac{z L \mathfrak{b}^2 \mathcal{F}}{4\kappa r_0^{z-2}\eta}
			+\frac{z^2 L \mathfrak{b}^2 \mathcal{F}^2}{4\kappa r_0^{z-2}\eta(z \mathcal{F}-4\eta \mathcal{F}')}
		\right]		\; .
\label{eq:chi}
\eea
Using the specific heat at constant volume,
\beq
\frac{c_{\mathcal{V}}}{T}
	=	\left.\frac{\partial\mathfrak{s}}{\partial T}\right|_{\mathfrak{b},\mathfrak{q}}	
	=	\frac{16 \pi^2}{\kappa L r_0^{2-z} \left(z \mathcal{F} - 4 \eta \mathcal{F}' \right)}	\; ,
\label{eq:c_V}
\eeq
as well as \eqnref{eq:Entropy} and \eqnref{eq:def_F}, \eqnref{eq:chi} could also be written as
\beq
\chi	=	\frac{(z\mathfrak{b}^2+8\kappa^2\mathfrak{q}^2)}{2(\mathfrak{b}^2+4\kappa^2\mathfrak{q}^2)}\frac{\mathfrak{m}}{\mathfrak{b}}
				-\frac{z T \mathfrak{b}^2 (2 \mathfrak{s}-z c_{\mathcal{V}}T)}{4(\mathfrak{b}^2+4\kappa^2\mathfrak{q}^2)^2}	\; .
\eeq
As a consequence, in the limit of vanishing temperature, assuming $\mathfrak{s}$ and $c_{\mathcal{V}}T$ do not diverge in this limit\footnote{Numerical investigations in \cite{Brynjolfsson:2010rx} suggest that they remain finite.},
\beq
\chi \Bigr|_{T=0}	=	\frac{(z\mathfrak{b}^2+8\kappa^2\mathfrak{q}^2)}{2(\mathfrak{b}^2+4\kappa^2\mathfrak{q}^2)}\frac{\mathfrak{m}}{\mathfrak{b}}	\; ,
\label{eq:chi_at_zero_T}
\eeq
and for vanishing magnetic field, 
\beq
\chi \Bigr|_{\mathfrak{b}=0}	=	\lim_{\mathfrak{b}\to 0}\frac{\mathfrak{m}}{\mathfrak{b}}	=	-\frac{1}{4\kappa^2}\frac{\mu}{\mathfrak{q}}	\; .
\label{eq:chi_at_zero_b}
\eeq
For $z=1$ this identity can easily be checked for dyonic AdS black branes.
That it also holds for $z>1$ based on \eqnref{eq:d_Helmholtz} is a non-trivial result.

As was noted in \cite{Dehghani:2011tx}, for $1\leq z<2$, the value of $\nu$ from appendix \ref{app:exps} can be expressed as
\beq
\nu = \int_0^1 r^{z-1} f dr	\; .
\eeq
Thus, because $f>0$ outside the horizon, from \eqnref{eg:magnetisation} follows that $\mathfrak{m}$ has the opposite sign to $\mathfrak{b}$.
As this also implies that $\chi$ will be negative, at least in the limits of low temperature and magnetic field strength, this means that the system exhibits diamagnetic behavior.
In contrast, for  $2\leq z < 6$, an expression for $\nu$ is given by
\beq
\nu = \int_0^1 r^{z-1} (f-1) dr - 
\left\{\begin{array}{lll}
			0		& \quad	&	z=2	\; ,	\\
			\frac{1}{z-2}	&\quad	&	2<z< 6	\; .
		\end{array}\right.
\eeq
Here it is potentially possible to have a setup with $\mathfrak{m}$ and $\mathfrak{b}$ having the same sign and $\chi$ positive and thus modeling a paramagnetic material.
In fact, the known exact solution for $z=4$ (see appendix \ref{app:z=4_sol}) is such a case.

\subsection{Numerical results}

Though the main results of this paper are derived analytically, it is instructive to also have a numerical check of certain equalities.
As the qualitative features seemed to be rather indifferent to the particular value of $z$, numerical results are just presented for one value, $z=\frac{3}{2}$.

\begin{figure}[h]
\begin{center}
\includegraphics[width=.35\paperwidth]{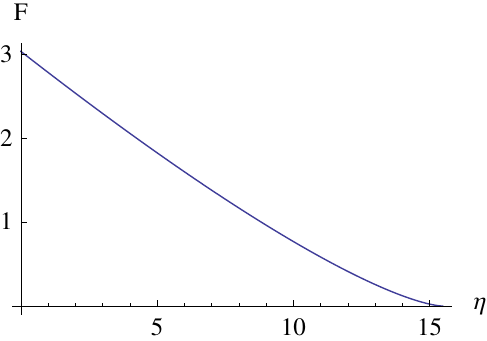}
\includegraphics[width=.35\paperwidth]{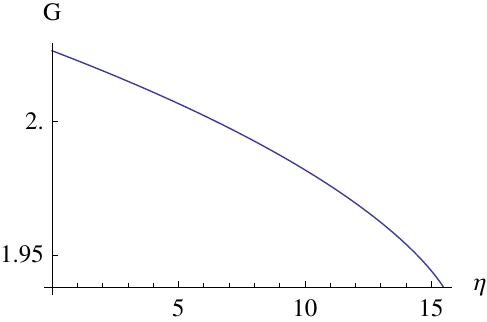}
\caption{Plot of the functions $\mathcal{F}(\eta)$ (left) and $\mathcal{G}(\eta)$ (right). $\mathcal{F}$ vanishes for $\eta=\frac{31}{2}$, indicating that the black brane becomes extremal for this value.}
\label{fig:FGplot}
\end{center}
\end{figure}

The information about thermodynamic quantities is encoded in the functions $\mathcal{F}$ and $\mathcal{G}$ \eqnref{eq:def_F} and \eqnref{eq:def_G}.
A plot of these functions can be seen in figure \ref{fig:FGplot}.
\label{sec:num}
The function $\mathcal{F}$ can be seen to vanish sublinearly at the critical value $\eta = -4\Lambda = \frac{31}{2}$.
As this happens at a finite value of $r_0$, this means that the black brane becomes extremal.
A more detailed numerical investigation of the behavior of the system when approaching criticality can be found in \cite{Brynjolfsson:2009ct}.

\begin{figure}[h]
\begin{center}
\includegraphics[width=.5\paperwidth]{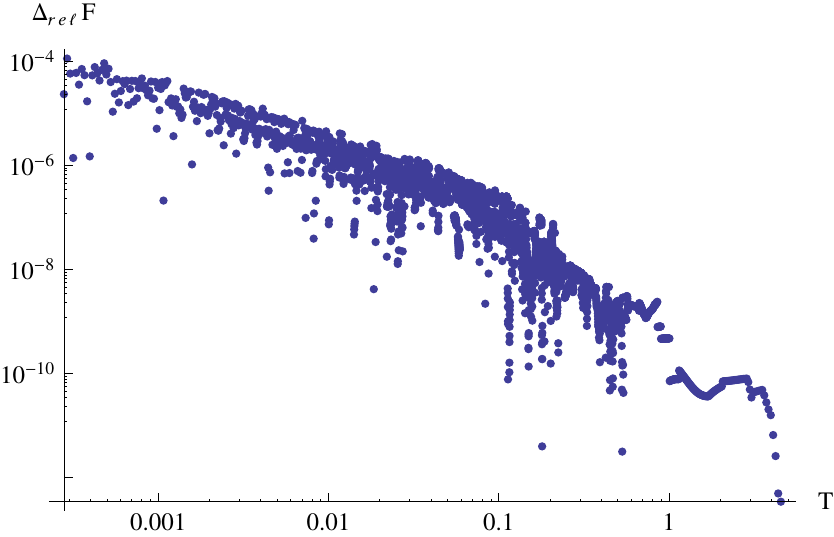}
\caption{The relative error $\Delta_{rel}\mathcal{F}$ versus temperature. The loss of precision for lower temperatures is due to the numerical function values become less precise in that region and due to an accumulation of numerical errors from the integration.}
\label{fig:delta_F}
\end{center}
\end{figure}

The aim now is to check \eqnref{eq:FGcond}.
To do this, define the functional 
\beq
\tilde{\mathcal{F}} = \mathcal{F}_0 - \eta \mathcal{G} + \frac{z+2}{4}\int_0^\eta \mathcal{G}	\; ,
\label{eq:Ftilde}
\eeq
which is the general solution of \eqnref{eq:FGcond} for given $\mathcal{G}$, and fix $\mathcal{F}_0$ such that $\tilde{\mathcal{F}}$ and $\mathcal{F}$ coincide at some value of $\eta$.
Then \eqnref{eq:Ftilde} can be compared with the numerical value of $\mathcal{F}$ at other values of $\eta$.
The relative error
\beq
\Delta_{rel}\mathcal{F} = 2\left|\frac{\mathcal{F}-\tilde{\mathcal{F}}}{\mathcal{F}+\tilde{\mathcal{F}}}\right|
\label{eq:delta_F}
\eeq
is plotted in figure \ref{fig:delta_F}.
For a better comparison later plots, $\eta$ has been translated back into a value of temperature, normalized by the temperature at $\mathfrak{q}=\frac{1}{2\kappa}$.
For temperatures of $O(1)$ and higher, the deviation can be seen to be lower than ten significant digits.
When the temperature is lowered, the deviation increases.
This can be attributed on the one hand to numerical values of $\mathcal{F}$ and $\mathcal{G}$ having lower precision at low temperatures and on the other hand to the accumulation of numerical errors when integrating \eqnref{eq:Ftilde}.

\begin{figure}[h]
\begin{center}
\includegraphics[width=.5\paperwidth]{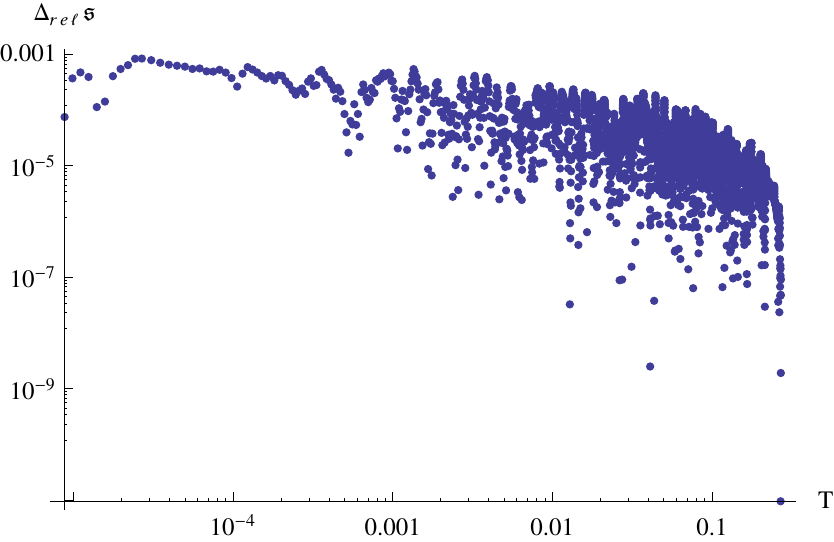}
\caption{The relative error $\Delta\mathfrak{s}$ versus temperature. The increase for lower values of temperature is due to the lower precision of the numerical data in that region.}
\label{fig:delta_s}
\end{center}
\end{figure}

As \eqnref{eq:FGcond} was shown to be equivalent to \eqnref{eq:d_Helmholtz}, the above results give a good indication that the latter is indeed satisfied.
It is however also be possible to make a more direct check.
For this purpose,
the relative error
\beq
\Delta_{rel}\mathfrak{s} = 2\left|\frac{\left.\frac{\partial\mathfrak{a}}{\partial T}\right|_{\mathfrak{b},\mathfrak{q}}+\mathfrak{s}}{\left.\frac{\partial\mathfrak{a}}{\partial T}\right|_{\mathfrak{b},\mathfrak{q}}-\mathfrak{s}}\right|
\label{eq:delta_s}
\eeq
is plotted in figure \ref{fig:delta_s}.
Results here are less precise than for \eqnref{eq:delta_F}.
This is mainly due to the fact that a numerical estimate of a derivative is in general more susceptible to the precision of the input data than the estimate of an integral.
Nevertheless, the numerics show an agreement to at least three significant digits - and even better agreement for higher temperatures, where the precision is better.

\begin{figure}[h]
\begin{center}
\includegraphics[width=.5\paperwidth]{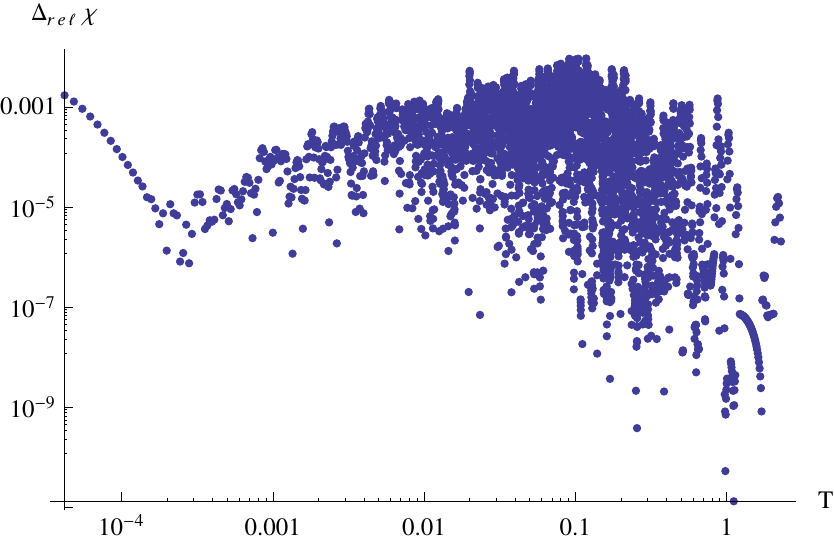}
\caption{The relative error $\Delta_{rel}\chi$ versus temperature for $\mathfrak{b}=0$, normalized with the values at $\mathfrak{q}=\frac{1}{2\kappa}$. As in the other plots, the error has a tendency to increases with lower values of $T$ due to precision issues in that region.}
\label{fig:delta_chi}
\end{center}
\end{figure}

This section concludes with a numerical check of \eqnref{eq:chi_at_zero_b}, which also was a consequence of \eqnref{eq:FGcond}.
This identity allows to compare a second derivative of $\mathfrak{a}$ with quantities that can be read off from the asymptotics.
Figure \ref{fig:delta_chi} shows the relative error
\beq
\Delta_{rel}\chi = 2\left|\frac{\left.\frac{\partial^2\mathfrak{a}}{\partial \mathfrak{b}^2}\right|_{T,\mathfrak{q}}-\frac{1}{4\kappa^2}\frac{\mu}{\mathfrak{q}}}{\left.\frac{\partial^2\mathfrak{a}}{\partial \mathfrak{b}^2}\right|_{T,\mathfrak{q}}+\frac{1}{4\kappa^2}\frac{\mu}{\mathfrak{q}}}\right|
\eeq
at vanishing magnetic field.
Also here an agreement of about three significant digits or better can be seen.

The results presented here were as far as it was possible to go using numerics with reasonable computation time.
The general trend was that the deviations presented above decreased when the precision was increased.
It stands to reason that computations done with an even higher number of significant digits would further improve the numerical results.

\section{Conclusions}
In this paper, several results on of dyonic Lifshitz black branes were established.

Though the renormalization of the action presented in this paper is still work in progress, preliminary results for black branes with $d=2$ and $z<6$ can be obtained.
The task of deriving an expression for general values of $d$, higher values of $z$ and even a curved horizon seems straightforward, however, an implementation is expected to become more complicated when expansions up to higher order need to be taken into account.

A further result would be the evidence that this renormalization procedure gives expressions for the Helmholtz free energy and internal energy that are in agreement with standard thermodynamic relations.
This bolsters the case for dyonic Lifshitz black branes as candidates for a holographic description of phenomena involving magnetism in quantum critical systems with a non-trivial dynamical exponent.

Finally, the magnetization and susceptibility of the dual theories were worked out using the gravitational description.
For $1\leq z <2$, the low temperature limit is always diamagnetic, whereas paramagnetism can occur for $2\leq z <6$.
The only known exact solution for $z=4$ happens to be of the paramagnetic type, but it remains an open question whether paramagnetism is the rule in this region of parameter space.

\section*{Acknowledgments}

I would like to thank L\'arus Thorlacius and Paata Kakashvili for interesting discussions and remarks during the process of writing this paper.
This work was supported by the Icelandic Research Fund, the University of Iceland Research Fund and the Eimskip Research Fund at the University of Iceland.

\appendix

\section{Asymptotic Expansions}
\label{app:exps}

An expansion for generic $z$ for asymptotically Lifshitz solutions of \eqnref{eq:bg_eqns_f}-\eqnref{eq:bg_eqns_phi} is as follows,
\bea
f	&=&	f_\infty \Bigl[
			1
			-\frac{ 2 \sqrt{z-1} (z_1-2) \Xi _1}{z (-2+2 z+z_1)} r^{z_1}
			- \frac{ \sqrt{z-1} (z_2-2) \Xi _2}{z (-2+2 z+z_2)} r^{z_2}	\nonumber	\\
	& &	-\left[\frac{4 \sqrt{z-1} \mathcal{M}}{2+z}+\frac{4 \left(z^2-2 z+2\right) \Xi _1 \Xi _2}{z \left(z^2 + 3z + 2\right)}\right]r^{2+z}	\nonumber	\\
	& &		- \frac{(z-1)(z-4) ({\cal B}^2+{\cal Q}^2)}{8 (z-2)^2 (z+1)} r^4
			+ \ldots \Bigr]	\label{eq:f_asymptotics}	\; ,	\\
g	&=&	
		1
		+\frac{\sqrt{z-1} z_1 \Xi _1}{z (-2+2 z+z_1)} r^{z_1}
		+\frac{\sqrt{z-1} z_2 \Xi _2}{z (-2+2 z+z_2)} r^{z_2}	\nonumber	\\
	& &	+\left[\sqrt{z-1} \mathcal{M}+\frac{\left(2z^3 - 5z^2 + 3z -2\right) \Xi _1 \Xi _2}{2 z^2 (z+1)}\right]r^{2+z}	\nonumber	\\
	& &	-\frac{(2 z-3) \left({\cal B}^2+{\cal Q}^2\right)}{4 (z-2)^2 (z+1)} r^4
		+ \ldots	\; ,
	\label{eq:g_asymptotics}		\\
a	&=&	
		\sqrt{z-1}
		+\frac{z_1 (z_1-2) \Xi _1}{z (-2+2 z+z_1)} r^{z_1}
		+\frac{z_2 (z_2-2) \Xi _2}{z (-2+2 z+z_2)} r^{z_2}	\nonumber	\\
	& &	+\frac{(z-4) \sqrt{z-1} \left({\cal B}^2+{\cal Q}^2\right)}{4 (z-2)^2 (z+1)}r^4		
		+2 {\cal M} r^{2+z}
		+ \ldots	\; ,	\label{eq:a_asymptotics} \\
b	&=&	
		\sqrt{z-1}
		-\frac{2 r^{z_1} z_1 \Xi _1}{z (-2+2 z+z_1)}	
		-\frac{2 r^{z_2} z_2 \Xi _2}{z (-2+2 z+z_2)}	\nonumber	\\
	& &	-\frac{(z-4) \sqrt{z-1} \left({\cal B}^2+{\cal Q}^2\right)}{4 (z-2)^2 (z+1)}r^4		
		-\frac{4 {\cal M}}{z}r^{2+z}
		+ \ldots	\; ,	\label{eq:b_asymptotics}	\\
\phi	&=&	{\cal Q}\left[- r^z \nu
		+ \frac{1}{2-z} r^2
		+ \ldots\right]	\; .		\label{eq:phi_asymptotics}	
\eea
where $\ldots$ indicate descendants of the previous listed modes.
The Exponents $z_1$ and $z_2$ are given by
\begin{eqnarray}
z_1	&=&	\frac{1}{2}\left[z+2 - \sqrt{(2+z)^2 + 8(z-1)(z-2)} \right]	\; ,	\label{eq:z_1}	\\
z_2	&=&	\frac{1}{2}\left[z+2 + \sqrt{(2+z)^2 + 8(z-1)(z-2)} \right]	\; .	\label{eq:z_2}
\end{eqnarray}
For $z>2$, the exponent $z_1>0$ and thus, in order to be renormalizable, the solution must have $\Xi_1=0$.
For the marginal case $z=2$ with $z_1=0$ and $z_2=4$, there are logarithmic modes occurring.
After discarding a growing mode, the expansions \eqnref{eq:f_asymptotics}-\eqnref{eq:phi_asymptotics} are modified to
\bea
f	&=&	1
		+r^4	\Biggl[
				\frac{32\mathcal{M}-32\,\Xi_2-\mathcal{B}^2-\mathcal{Q}^2}{96} 
				+\frac{8\mathcal{M}+\mathcal{B}^2+\mathcal{Q}^2}{16} \ln r	\nonumber	\\
	& &	\qquad\qquad	+\frac{\mathcal{B}^2+\mathcal{Q}^2}{16} (\ln r)^2
			\Biggr]
		+ \ldots	\; ,	\\
g	&=&	1
		+r^4	\Biggl[
				\frac{28 \mathcal{M}+32\,\Xi_2 + \mathcal{B}^2+\mathcal{Q}^2}{96}
				+\frac{-16\mathcal{M}+3(\mathcal{B}^2+\mathcal{Q}^2)}{32} \ln r	\nonumber	\\
	& &	\qquad\qquad	-\frac{\mathcal{B}^2+\mathcal{Q}^2}{16} (\ln r)^2 \Biggr]
		+ \ldots	\; ,	\\
a	&=&	1+r^4	\Biggl[
				-\frac{88 \mathcal{M}-64\,\Xi_2 +\mathcal{B}^2+\mathcal{Q}^2}{96}
				-\frac{16\mathcal{M} + 3(\mathcal{B}^2+\mathcal{Q}^2)}{16} \ln r	\nonumber	\\
	& &	\qquad\qquad	-\frac{\mathcal{B}^2+\mathcal{Q}^2}{8} (\ln r)^2
			\Biggr]
		+ \ldots	\; ,	\\
b	&=&	1+r^4	\Biggl[
				\frac{20 \mathcal{M} -32\,\Xi_2 -\mathcal{B}^2-\mathcal{Q}^2}{48}
				+\frac{16\mathcal{M} + \mathcal{B}^2+\mathcal{Q}^2}{16} \ln r	\nonumber	\\
	& &	\qquad\qquad	+\frac{\mathcal{B}^2+\mathcal{Q}^2}{8} (\ln r)^2
			\Biggr]
		+ \ldots	\; ,	\\
\phi	&=&	\mathcal{Q} \left[ r^2 (-\nu + \ln r\,)
		+ \ldots	\right]	\; .	
\eea

\section{Exact solutions}
\label{app:exact_solutions}

\subsection{z=1 : dyonic AdS black holes}
\label{app:dyonic_AdSBH}

The solution of a dyonic black brane with a horizon at $r=1$ is given by
\beq
f=1,
g^2=1 - \left(1 + \frac{\rho_0^2 + B_0^2}{4} \right) r^3 + \frac{\rho_0^2 + B_0^2}{4} r^4,
a=b=0,
\phi=\frac{\rho_0}{g^2} \left(- r + r^2 \right)	\; .
\eeq
From this follow the thermodynamic quantities
\bea
\mathfrak{q}	&=&	\frac{\rho_0}{2\kappa L^2 r_0^2}	\; ,	\\
\mu		&=&	\frac{\rho_0}{L r_0}	\; ,	\\
\mathfrak{b}	&=&	\frac{B_0}{L^2 r_0^2}	\; ,	\\
\mathfrak{m}	&=&	-\frac{B_0}{2\kappa L r_0}	\; ,	\\
T		&=&	\frac{12-\rho_0^2-B_0^2}{16 \pi L r_0}	\; ,	\\
\mathfrak{s}	&=&	\frac{2 \pi}{\kappa L^2 r_0^2}	\; ,	\\
\mathfrak{e}	&=&	\frac{4 + \rho_0^2 + B_0^2}{2 \kappa L^3 r_0^3}	\; ,	\\
\mathfrak{a}	&=&	\frac{-4 + 3\rho_0^2 + 3B_0^2}{8 \kappa L^3 r_0^3}	\; ,	\\
\frac{c_{\mathcal{V}}}{T}	&=&	\frac{64 \pi ^2}{3 \kappa L r_0 \left(4 + \rho_0^2 + B_0^2\right)}	\; ,	\\
\chi		&=&	-\frac{L r_0 \left(12 + 3\rho_0^2 + B_0^2\right)}{6\kappa  \left(4 + \rho_0^2 + B_0^2\right)}	\; .
\eea

\subsection{z=4}
\label{app:z=4_sol}
For $z=4$, there is the special solution\footnote{This basically is the solution presented in \cite{Brynjolfsson:2009ct}, rotated on the $B_0\rho_0$ plane.}
\beq
f = 1,
g^2 = 1-r^4,
a = \sqrt{3},
b = \sqrt{3},
\Phi = -\frac{\rho_0 r^2}{2 \left(r^2+1\right)},
\eeq
which solves \eqnref{eq:bg_eqns_f}-\eqnref{eq:bg_eqns_phi} provided $\eta=B_0^2+\rho_0^2=8$.
As it is just an isolated solution at a single value of $\eta$, the thermodynamic quantities that involve differentiation can not be calculated.
The ones obtainable are
\bea
\mathfrak{q}	&=&	\frac{\rho_0}{2\kappa L^2 r_0^2}	\; ,	\\
\mu		&=&	-\frac{\rho_0}{2 L r_0^4}	\; ,	\\
\mathfrak{b}	&=&	\frac{B_0}{L^2 r_0^2}	\; ,	\\
\mathfrak{m}	&=&	\frac{B_0}{4\kappa L^2 r_0^4}	\; ,	\\
T		&=&	\frac{1}{\pi L r_0^4}	\; ,	\\
\mathfrak{s}	&=&	\frac{2 \pi}{\kappa L^2 r_0^2}	\; ,	\\
\mathfrak{e}	&=&	0	\; ,	\\
\mathfrak{a}	&=&	\frac{-2}{\kappa L^3 r_0^6}	\; .
\eea
It might be worth noting that in this solution $\mathfrak{q}$ and $\mu$ have opposite sign and the internal energy is vanishing.

\newpage

\bigskip
\bibliography{DLBH}

\providecommand{\href}[2]{#2}\begingroup\raggedright\begin{thebibliography}{10}

\bibitem{Maldacena:1997re}
J.~M. Maldacena, {\it {The large N limit of superconformal field theories and
  supergravity}},  {\em Adv. Theor. Math. Phys.} {\bf 2} (1998) 231--252,
  [\href{http://xxx.lanl.gov/abs/hep-th/9711200}{{\tt hep-th/9711200}}].

\bibitem{Kachru:2008yh}
S.~Kachru, X.~Liu, and M.~Mulligan, {\it {Gravity Duals of Lifshitz-like Fixed
  Points}},  {\em Phys. Rev.} {\bf D78} (2008) 106005,
  [\href{http://xxx.lanl.gov/abs/0808.1725}{{\tt arXiv:0808.1725}}].

\bibitem{2005Natur.433..226C}
P.~{Coleman} and A.~J. {Schofield}, {\it {Quantum criticality}},  {\em NATURE}
  {\bf 433} (Jan., 2005) 226--229,
  [\href{http://xxx.lanl.gov/abs/cond-mat/0503002}{{\tt cond-mat/0503002}}].

\bibitem{Hartnoll:2009sz}
S.~A. Hartnoll, {\it {Lectures on holographic methods for condensed matter
  physics}},  {\em Class. Quant. Grav.} {\bf 26} (2009) 224002,
  [\href{http://xxx.lanl.gov/abs/0903.3246}{{\tt arXiv:0903.3246}}].

\bibitem{McGreevy:2009xe}
J.~McGreevy, {\it {Holographic duality with a view toward many-body physics}},
  {\em Adv. High Energy Phys.} {\bf 2010} (2010) 723105,
  [\href{http://xxx.lanl.gov/abs/0909.0518}{{\tt arXiv:0909.0518}}].

\bibitem{Sachdev:2010ch}
S.~Sachdev, {\it {Condensed matter and AdS/CFT}},
  \href{http://xxx.lanl.gov/abs/1002.2947}{{\tt arXiv:1002.2947}}.

\bibitem{Hartnoll:2007ai}
S.~A. Hartnoll and P.~Kovtun, {\it {Hall conductivity from dyonic black
  holes}},  {\em Phys. Rev.} {\bf D76} (2007) 066001,
  [\href{http://xxx.lanl.gov/abs/0704.1160}{{\tt arXiv:0704.1160}}].

\bibitem{Taylor:2008tg}
M.~Taylor, {\it {Non-relativistic holography}},
  \href{http://xxx.lanl.gov/abs/0812.0530}{{\tt arXiv:0812.0530}}.

\bibitem{Bertoldi:2009vn}
G.~Bertoldi, B.~A. Burrington, and A.~Peet, {\it {Black Holes in asymptotically
  Lifshitz spacetimes with arbitrary critical exponent}},  {\em Phys. Rev.}
  {\bf D80} (2009) 126003, [\href{http://xxx.lanl.gov/abs/0905.3183}{{\tt
  arXiv:0905.3183}}].

\bibitem{Brynjolfsson:2009ct}
E.~J. Brynjolfsson, U.~H. Danielsson, L.~Thorlacius, and T.~Zingg, {\it
  {Holographic Superconductors with Lifshitz Scaling}},  {\em J. Phys.} {\bf
  A43} (2010) 065401, [\href{http://xxx.lanl.gov/abs/0908.2611}{{\tt
  arXiv:0908.2611}}].

\bibitem{Ross:2009ar}
S.~F. Ross and O.~Saremi, {\it {Holographic stress tensor for non-relativistic
  theories}},  {\em JHEP} {\bf 09} (2009) 009,
  [\href{http://xxx.lanl.gov/abs/0907.1846}{{\tt arXiv:0907.1846}}].

\bibitem{Dehghani:2011tx}
M.~H. Dehghani, R.~B. Mann, and R.~Pourhasan, {\it {Charged Lifshitz Black
  Holes}},  \href{http://xxx.lanl.gov/abs/1102.0578}{{\tt arXiv:1102.0578}}.

\bibitem{Brynjolfsson:2010rx}
E.~J. Brynjolfsson, U.~H. Danielsson, L.~Thorlacius, and T.~Zingg, {\it {Black
  Hole Thermodynamics and Heavy Fermion Metals}},  {\em JHEP} {\bf 08} (2010)
  027, [\href{http://xxx.lanl.gov/abs/1003.5361}{{\tt arXiv:1003.5361}}].

\bibitem{Brynjolfsson:2010mk}
E.~J. Brynjolfsson, U.~H. Danielsson, L.~Thorlacius, and T.~Zingg, {\it
  {Holographic models with anisotropic scaling}},
  \href{http://xxx.lanl.gov/abs/1004.5566}{{\tt arXiv:1004.5566}}.

\bibitem{Balasubramanian:1999re}
V.~Balasubramanian and P.~Kraus, {\it {A stress tensor for anti-de Sitter
  gravity}},  {\em Commun. Math. Phys.} {\bf 208} (1999) 413--428,
  [\href{http://xxx.lanl.gov/abs/hep-th/9902121}{{\tt hep-th/9902121}}].

\bibitem{Hartnoll:2009ns}
S.~A. Hartnoll, J.~Polchinski, E.~Silverstein, and D.~Tong, {\it {Towards
  strange metallic holography}},  {\em JHEP} {\bf 04} (2010) 120,
  [\href{http://xxx.lanl.gov/abs/0912.1061}{{\tt arXiv:0912.1061}}].

\bibitem{FeffermanGraham}
C.~Fefferman and C.~Robin~Graham, {\it {Conformal Invariants}},  in {\em {Elie
  Cartan et les math\'ematiques d'aujourd'hui}}, {Ast\'erisque}, pp.~95--116,
  {Soci\'et\'e Math\'ematique de France, Paris}, June, 1985.
\newblock {hors s\'erie}.

\bibitem{TBA}
T.~Zingg, {\it {in preparation}}, .

\bibitem{Bertoldi:2009dt}
G.~Bertoldi, B.~A. Burrington, and A.~W. Peet, {\it {Thermodynamics of black
  branes in asymptotically Lifshitz spacetimes}},  {\em Phys. Rev.} {\bf D80}
  (2009) 126004, [\href{http://xxx.lanl.gov/abs/0907.4755}{{\tt
  arXiv:0907.4755}}].

\end{thebibliography}\endgroup
\bibliographystyle{JHEP}

\end{document}